\documentclass[twocolumn,showpacs,prb]{revtex4}% Physical Review B

\usepackage{graphicx}% Include figure files
\usepackage{dcolumn}% Align table columns on decimal point
\usepackage{bm}% bold math

\begin{document}

% title for ArXiv:
\title{Electric quadrupole and magnetic dipole coupling in plasmonic nanoparticle arrays}
% Another title - it will be for the paper:
%\title{Metasurfaces with electric quadrupole and magnetic dipole resonant coupling}% Force line breaks with \\

\author{Viktoriia E. Babicheva}
\affiliation{College of Optical Sciences, University of Arizona, Tucson, AZ, USA}
\email{vbab.dtu@gmail.com}

\author{Andrey B. Evlyukhin}
\affiliation{Laser Zentrum Hannover e.V., Hollerithallee 8,
	D-30419 Hannover, Germany}
\affiliation{ITMO University, St. Petersburg 197101, Russia} 

\date{\today}% It is always \today, today,
% but any date may be explicitly specified

\date{\today}% It is always \today, today,
% but any date may be explicitly specified

\begin{abstract}
Collective resonances in plasmonic nanoparticle arrays with electric dipole moment oriented along the lattice wave propagation are theoretically investigated. The role of electric quadrupole (EQ) and magnetic dipole (MD) moments of gold nanoparticles in the resonant features of the arrays is analyzed. We perform both semi-analytical calculations of coupled multipole equations and rigorous numerical simulations varying contributions of the electric and magnetic multipoles by changing particle size and shape (spheres and disks). The arrays in homogeneous and non-homogeneous environments are considered. We find that even very weak non-resonant EQ and MD moments of a single particle are significantly enhanced in the periodic lattice at the wavelength of collective (lattice) resonance excitation. Importantly, we show that in the infinite arrays, the EQ and MD moments of nanoparticles are coupled and affect each other resonant contributions. We also demonstrate that at the lattice-resonance wavelength, the enhanced EQ and MD moments have contributions to reflection comparable to the dipole one resulting in a significant decrease of reflection and providing the satisfaction of the generalized Kerker condition for reflection suppression.
\end{abstract}
\maketitle

\section{Introduction}
Recently, periodic arrays of plasmonic nanoparticles have gained a lot of interest as they can produce narrow collective plasmon resonances in light transmission spectra because of the nanoparticle coupling in the array [1-15]. These resonances appear at the wavelength of scattered light close to the wavelength corresponding to the Rayleigh anomaly. The strength and width of the resonances are characterized by the spectral distance between the closest Rayleigh anomaly and the single-particle plasmon resonance. The wavelengths corresponding to the Rayleigh anomaly are determined by the array periods, and the wavelength of the single-particle plasmon resonance is determined by nanoparticle size and shape. The spectral sensitivity to the environmental properties of collective resonances makes them very attractive for sensing applications [16]. Resonant properties can also be utilized to enhance the performance of light sources [13], nanolasers [17], modulators [18], solar cells [19,20], and narrow-band photodetectors [21], and the enabled functionalities are discussed in detail in the recent review [22].  

Various nano-scatterers were shown to enhance lattice resonances for orthogonal and parallel plasmonic-photonic coupling [23-26], the dipole coupling of several particles in the cell has been studied in [27], and multipolar interactions in the surface-lattice resonances in two-dimensional arrays of spheres has been analyzed in [28,29].
In polarization where an electric field of the incident wave is perpendicular to direction with a periodicity of interest, resonances are attributed to dipole-dipole coupling between nanoparticles, and they are well studied. No resonant feature can appear in dipole approximation, homogeneous environment, and polarization where the electric field of the incident wave is parallel to the direction with a periodicity of interest [3]. Thus, recently observed lattice resonance in parallel polarization [30] cannot be attributed to the pure dipole-dipole coupling between nanoparticles and more studies are necessary to understand properties of such structures. In a more general case when excitations of magnetic multipoles are possible, e.g. in high-index dielectric nanoparticles, the lattice resonances appear in parallel polarization even for the dipole approximation, that is when only magnetic dipole (MD) resonances of each nanoparticle are excited [9,10]. In this work, we restrict ourselves to the case of metal nanoparticles, where MD resonances are weak and do not provide resonant feature without coupling to other multipoles.

According to the first Kerker condition [31], if electric and magnetic polarizabilities of a nanoparticle are equal each other in magnitude and phase, light scattering from this nanoparticle is suppressed in the backward direction (Kerker effect). For silicon spherical nanoparticle array, electric dipole (ED) and MD resonances do not overlap, and only non-resonant Kerker effect is possible [9,32-36], but starting with the pioneering work [37], a number of studies suggest obtaining spectral overlap of ED and MD resonances using all-dielectric nanoparticles with a more complex shape, such as disks, cubes, cones, or pyramids [34,38-41]. Furthermore, Kerker conditions have been generalized to include contributions of different multipole modes and achieve directional scattering with plasmonic [42,43], dielectric [44], or hyperbolic-material particles [45,46]. Recently, it has been shown that resonant Kerker effect is possible upon overlap of ED and MD lattice resonances excited in the periodic array of either silicon or core-shell nanoparticles, and spectral position of the resonant directional scattering can be controlled by the lattice size [10,47].

In this work, we study lattice resonances in periodic arrays of plasmonic nanoparticles, and we analyze different shapes (spheres and disks) and sizes of the nanoparticles that allows varying contributions of ED- and electric quadrupole (EQ) moments as well as different refractive indices of the surrounding environment (Fig. 1). Small plasmonic nanoparticles (with the radius being smaller than 60 nm) have only ED resonant contribution to the scattering in the visible spectral range, and the contributions of their EQ and MD moments to the scattering are negligibly small [8]. However, we show that even in the case of the homogeneous environment, small nanoparticles with small EQ and MD moments enable excitation of well pronounce lattice resonance due to EQ and MD coupling between nanoparticles in the array. Moreover, overlapping EQ and MD resonances with the broad ED response outside its lattice resonance, one can observe a suppression of total reflectance. We show that this decrease of reflectance is achieved because of Kerker condition of the directional scattering is satisfied. We derive its generalized form which takes into account ED, EQ, MD, and their coupling in the lattice.

\begin{figure}
	\begin{center}
		\includegraphics[scale=.100,width=20pc,height=18pc,keepaspectratio]{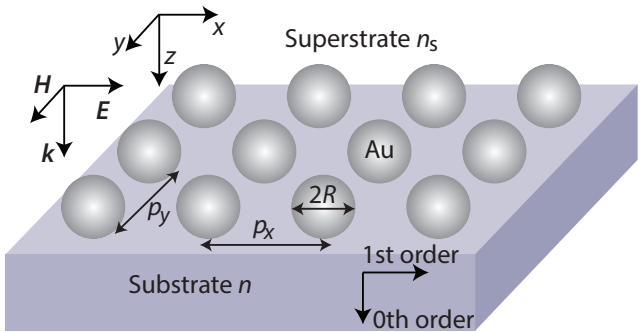}%{S_p_small}%{ReSumR120n1_45Rs660}% 
	\end{center}
	\caption{\label{F1}Schematic view of the nanosphere array with different refractive indices of substrate and superstrate ($n$ and $n_s$, respectively). Nanoparticles with radius $R$ are arranged into the array with the periods $p_x$ and $p_y$. Light incidence \textbf{k} is normal to the substrate, and electric field \textbf{E} is along the $x$-axis (referred as parallel polarization throughout the work).}
\end{figure}

Both sphere and disk shapes of nanoparticles are of interest. On the one hand, from a theoretical point of view, spheres are the most favorable, as they can be described by Mie theory, and exact solutions can be obtained by analytical approach only. On the other hand, for practical applications, the realization of a periodic array of disks is easier than an array of spheres, as disks can be fabricated by conventional optical lithography. Furthermore, in the present work, we demonstrate that for the non-homogeneous environment of nanoparticle array (substrate and superstrate are of different indices), electromagnetic coupling between nanoparticles causes a band of transparency in the transmission spectrum. We analyze the properties of the nanoparticle array on the substrate with different refractive indices and show that redistribution of energy in different diffraction orders causes peaks and dips in transmission and reflection spectra. From the practical point of view, we consider the small difference in indices of substrate and superstrate. In experiments, fabricated nanostructures are often covered with another medium to achieve index matching of the substrate [39,48] and obtain conditions close to the homogeneous environment in order to observe the strong lattice resonances.

The first part of the paper is devoted to the investigation of EQ-MD lattice resonances in homogeneous surrounding medium performed both numerically and analytically for sphere array. Through comparison of both calculations, we show that EQ-MD coupling is critical for correct calculations of the wavelength where generalized Kerker condition of directional scattering is satisfied. The second part considers an array of disks and besides homogeneous environment, we also study an influence of a small dielectric contrast between substrate and superstrate on transmission and reflection properties of the array.

\section{Methods and results}

\subsection{Nanoparticle array and irradiation conditions}
Similar to the Ref.[30], we study infinite gold nanoparticle array being on top of the substrate with index $n$ and covered with another material-superstrate-with refractive index $n_s = 1.47$ (fixed throughout the paper) in parallel polarization (Fig. 1). In this work, we perform numerical simulations (CST Microwave Studio) for both spheres and disks: sphere radius is varied, i.e. $R = 50$, 85, or 100 nm (Fig. 2); disk radius is fixed to $R_d = 85$ nm, and the height $H$ is either 50 or 100 nm. Gold permittivity is taken from experimental data [49]. 

\begin{figure}
	\begin{center}
		\includegraphics[scale=.100,width=20pc,height=18pc,keepaspectratio]{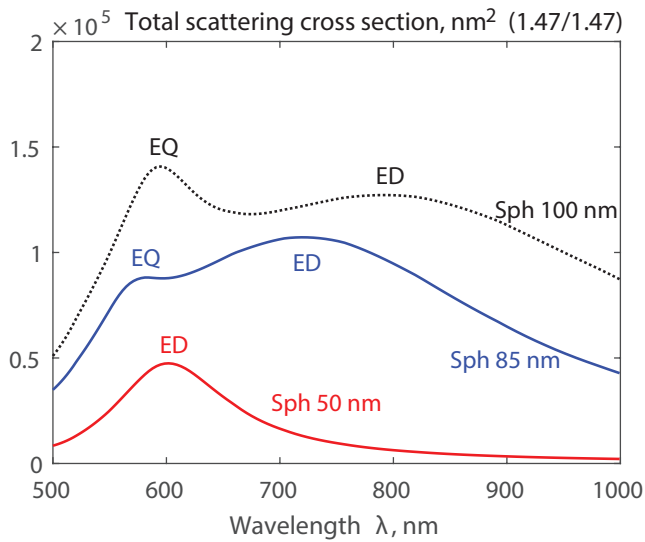}%{S_p_small}%{ReSumR120n1_45Rs660}% Here is how to import EPS art
	\end{center}
	\caption{\label{F2} Total scattering cross-section of the nanospheres under consideration, surrounded by a homogeneous environment with $n = n_s = 1.47$; numbers denote the radius of the spheres. ED denotes electric dipole, EQ denotes electric quadrupole, and no magnetic dipole resonances are excited for the single particle of such shape and dimensions. }
\end{figure}

We perform simulations for infinite nanoparticle array with periods $p_x = 510$ nm and $p_y = 250$ nm in the $x$- and $y$-directions, respectively. Arranging nanoparticles in the array with $p_x = 510$ nm in the medium with $n = 1.47$ is expected to cause lattice resonances spectrally close to the Rayleigh anomaly ($\lambda_{\text{RA}} = n\lambda_{\text{effRA}} = n p_x = 750$ nm). Simulations show that either for large enough nanoparticles or in the case of a small mismatch between the indices of substrate and superstrate, a strong resonant feature is observed (see below). We note that periodicity in the perpendicular direction, that is $p_y$, cannot cause dipole lattice resonances at the wavelength of interest because $\lambda_{\text{RA}} = n p_y = 368$ nm of the corresponding Rayleigh anomaly, and it is smaller than the wavelength of dipole resonance of the single particle [4]. Thus, the appearance of such pronounced band of transparency motivates us to study the effect in more detail.

Here we consider the only normal incidence of the waves, and the case of oblique incidence is beyond the scope of our current work. We analyze (i) the zeroth diffraction order of transmission/reflection, that is the propagation of the wave into the substrate/superstrate at normal direction; (ii) the first diffraction order of transmission/reflection; and (iii) full transmission/reflection of the array. We also calculate absorbance of nanoparticle array, i.e. the portion of energy absorbed inside nanoparticles.

\subsection{Numerical study of nanosphere array}

Let us consider nanoparticles in the homogeneous environment. Total scattering cross sections of single spherical nanoparticles, calculated in numerical simulations, are shown in Fig. 2. Multipole analysis of Ref. [8] allows associating the resonant maxima in Fig. 2 with resonant excitation of ED and EQ moments of the nanoparticles. For the single sphere with $R = 50$ nm, ED resonance is at the wavelength 600 nm, and there are no other multipole resonances as the particle is too small (Fig. 2, red line). Next, we calculate absorbance of the nanosphere array with $R = 50$ nm (Fig. 3a, red line) and see that the resonant feature at Rayleigh anomaly is not pronounced. For the single sphere with $R = 85$ nm, ED resonance is at wavelength 720 nm, and EQ resonance is at wavelength 580 nm (Fig. 2, blue line). From the comparison of the single particle scattering cross-section with the absorbance profiles of the array of such nanoparticles, one can see that in the array, EQ resonance appears at wavelength 540 nm. However, in contrast to small particles, in absorbance profile of particles with $R = 85$ nm, one can see a strong narrow peak close to $\lambda_{\text{RA}} = np_x = 750$ nm (Fig. 3a, blue line), which corresponds to the excitation of the collective (lattice) resonance. The same is observed for spheres with $R = 100$ nm, and the lattice resonance in absorption spectra is more pronounced than for spheres with $R = 85$ nm. 

\begin{figure*}
	\begin{center}
		\includegraphics[scale=.100,width=40pc,height=12pc,keepaspectratio]{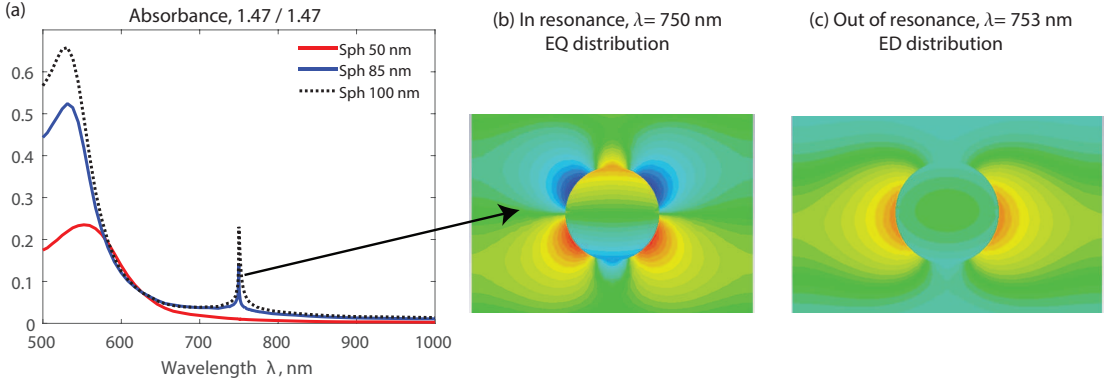}%{S_p_small}%{ReSumR120n1_45Rs660}% Here is how to import EPS art % 
	\end{center}
	\caption{\label{F3} (a) Absorbance of the nanosphere array in the homogeneous environment with $n = n_s = 1.47$. Periods are $p_x = 510$ nm and $p_y = 250$ nm. (b) In-resonance ($\lambda = 750$ nm) and (c) out-of-resonance ($\lambda = 753$ nm) field distributions for the array of nanospheres with $R = 85$ nm (one unit cell, side view). EQ field distribution is well pronounced for the resonant wavelength, and the field distribution drastically changes to ED with the small change of the wavelength. }
\end{figure*}

As it follows from the symmetry of the system, under normal incidence with electric field \textbf{E} along the $x$-axis, the dipole moment of nanoparticles is excited in the $x$-direction. In this case, neither dipole coupling normal to the array nor coupling in the direction of dipoles can contribute to lattice mode. In other words, in the homogeneous environment and dipole approximation, no resonant feature related to lattice mode can appear, and it is observed in numerical simulations of small particles, e.g. spheres with radius $R = 50$ nm (Fig. 3a). Yet another process dominates for larger particles, e.g. with $R = 85$ or 100 nm. Even for the homogeneous environment, one can see resonant features appear at $\lambda > \lambda_{\text{RA}} = np_x = 750$ nm. As one can see from the field distribution in Fig. 3b,c, these resonances correspond to the lattice modes existed with EQs. Comparing in-resonance ($\lambda = 750$ nm) and out-of-resonance ($\lambda = 753$ nm) field distributions for the array of nanospheres with $R = 85$ nm, one can see that EQ field distribution is well pronounced for the resonant wavelength, and the field drastically transforms to ED distribution upon the small spectral shift. At the wavelength of lattice resonance, MD resonances are also excited, but their field distribution does not have a pronounced pattern, and it is difficult to distinguish them from other multipoles.

The pronounced lattice resonances for the spheres with $R = 85$ or 100 nm cause strong resonant profiles in reflectance and transmittance spectra of the arrays (Fig. 4). For $R = 85$ nm, reflectance minimum close to zero is observed. For $R = 100$ nm, both reflectance maximum and minimum appear. It means that the excited EQ resonance is strong enough to provide destructive interference with ED and to suppress backscattering: at the wavelength of EQ lattice resonance, the reflection spectrum drops to zero which means that Kerker-like effect occurs. In the next subsection, we perform analytical calculations taking into account ED, EQ, and MD contributions and show that indeed, the observed effect results from the satisfying generalized Kerker condition with all three multipoles.

\begin{figure}
	\begin{center}
		\includegraphics[scale=.100,width=20pc,height=15pc,keepaspectratio]{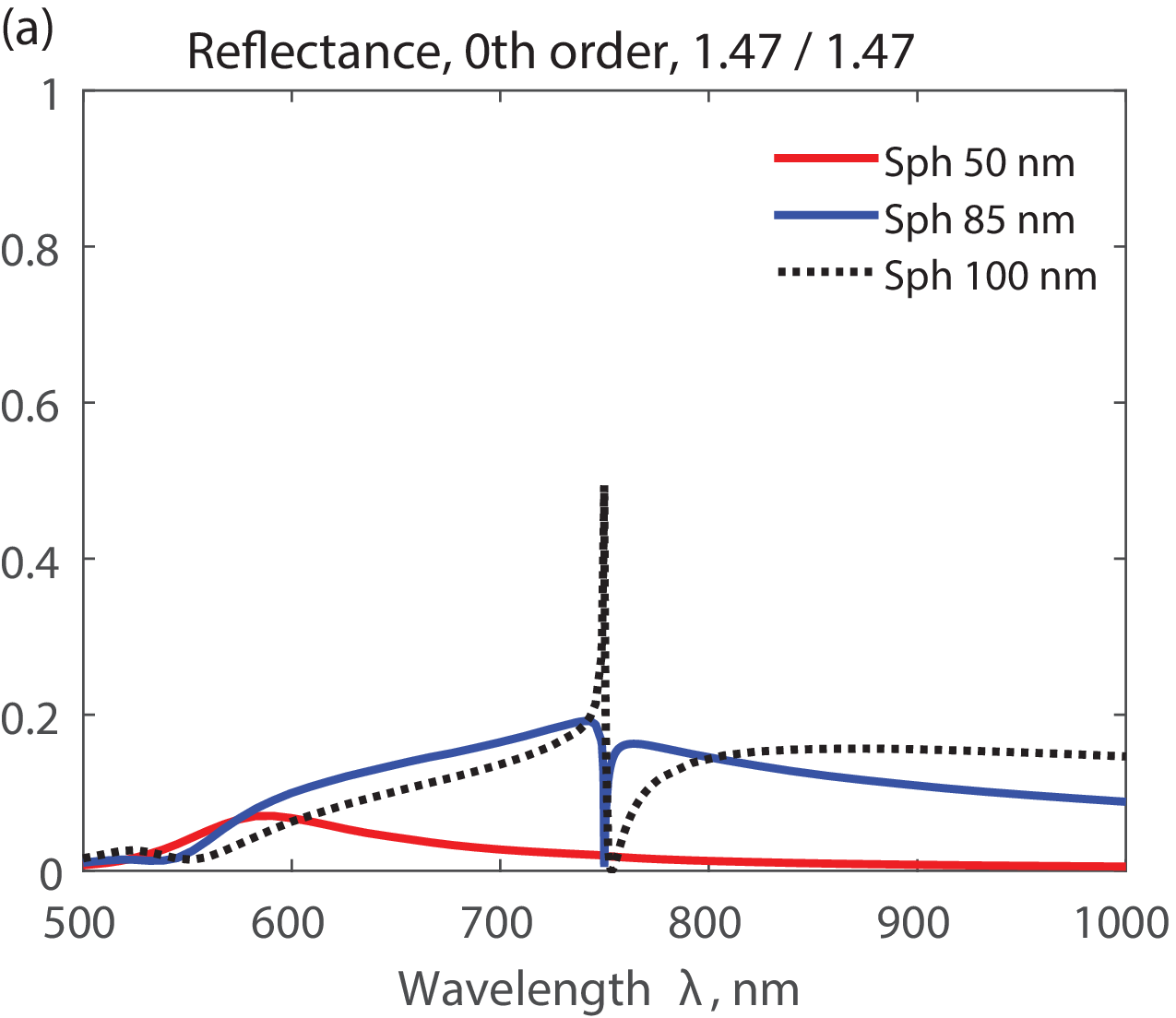}%{S_p_small}%{ReSumR120n1_45Rs660}% Here is how to import EPS art		
		
		\includegraphics[scale=.100,width=20pc,height=15pc,keepaspectratio]{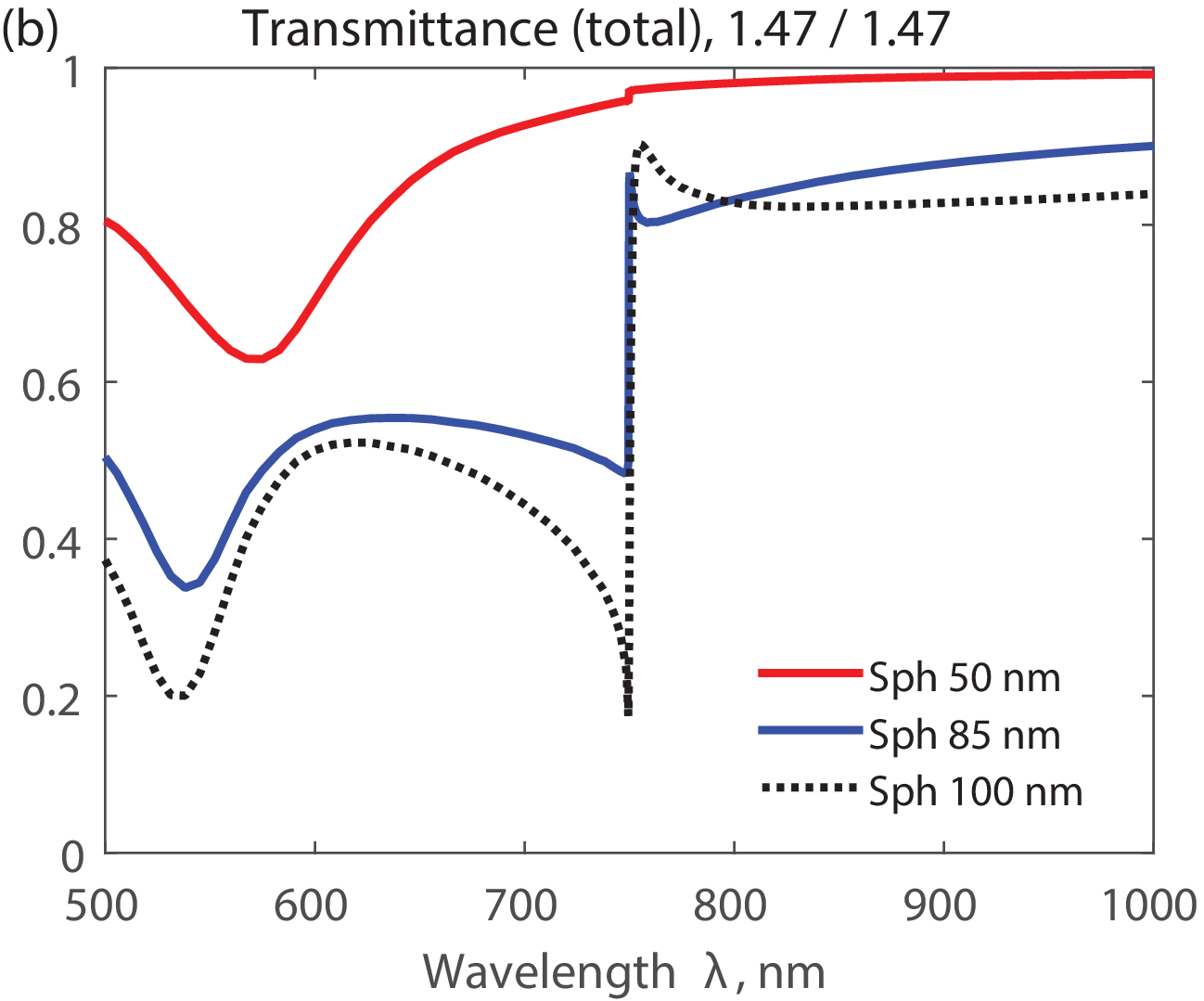}%
	\end{center}
	\caption{\label{F4} Lattice resonance in an array of nanospheres: (a) Zero-order reflectance and (b) total transmittance. Periods are $p_x = 510$ nm and $p_y = 250$ nm. See more simulation results in Supporting Information including the case of the non-homogeneous environment, when the refractive indices of the superstrate and substrate are $n_s = 1.47$ and $n = 1.5$, respectively (Figs. S1 and S2). }
\end{figure}

\subsection{Dipole-quadrupole model}

In the dipole-quadrupole approximation, the spherical particles are
considered as point electric and magnetic dipoles and electric quadrupoles with scalar dipolar and quadrupolar polarizabilities $\alpha_p$, $\alpha_m$, and $\alpha_q$,
respectively. The vectors of the electric dipole ${\bf p}^j$ and magnetic dipole ${\bf m}^j$ moments 
and the tensor of the electric quadrupole moment ${\hat Q}^j$ of arbitrary 
particle with number $j$ in the structure of $N$ particles are determined by the coupled equations (which are obtained the equations from [8,9])
\begin{eqnarray}\label{E_pmQ}
{\bf p}^j&=&\alpha_p\frac{k_0^2}{\varepsilon_0}\sum\limits_{l\neq
	j}^{N}[\hat G_{jl}^p{\bf p}^l +\frac{i}{ck_0}[{\bf g}_{jl}\times{\bf m}^l]+\hat G_{jl}^Q(\hat Q^l{\bf
	n}_{lj})]\nonumber\\
&&+\alpha_p{\bf E}_0({\bf
	r}_j)\:,\nonumber\\
{\bf m}^j&=&\alpha_m{k_0^2}\sum\limits_{l\neq
	j}^{N}[{\varepsilon_S}\hat G_{jl}^p{\bf m}^l +\frac{c}{ik_0}[{\bf{g}}_{jl}\times{\bf p}^l]\nonumber\\
&&+\frac{c}{ik_0}[{\bf q}_{jl}\times(\hat Q^l{\bf
	n}_{lj})]]+\alpha_m{\bf H}_0({\bf
	r}_j)\:,\nonumber\\
{\hat Q}^j&=&\frac{\alpha_q}{2}[\nabla{\bf E}_0({\bf r}_j)+{\bf
	E}_0({\bf
	r}_j)\nabla]\\
&+&\frac{\alpha_qk_0^2}{2\varepsilon_0}\sum\limits_{l\neq j}^{N}[
\nabla_j(\hat G_{jl}^p{\bf p}^l)+(\hat G_{jl}^p{\bf p}^l)\nabla_j]\nonumber\\
&+&\frac{\alpha_qk_0^2}{2\varepsilon_0}\sum\limits_{l\neq
	j}^{N}[\nabla_j(\hat G_{jl}^Q(\hat Q^l{\bf n}_{lj}))+(\hat
G_{jl}^Q(\hat Q^l{\bf n}_{lj}))\nabla_j]\nonumber\\
&+&\frac{\alpha_qk_0^2}{2\varepsilon_0}\frac{i}{ck_0}\sum\limits_{l\neq
	j}^{N}[\nabla_j[{\bf g}_{jl}\times{\bf m}^l]+[{\bf g}_{jl}\times{\bf m}^l]\nabla_j]\:,\nonumber
\end{eqnarray}
where $\nabla_j$ is the nabla operator with respect to ${\bf r}_j$, ${\bf E}_0({\bf r}_j)$ and ${\bf H}_0({\bf r}_j)$ are the incident fields at the
point of the particles with number $j$ (under condition of normal
incidence, which is considered in the paper, these fields are the
same for all particles in the structures), $k_0$ is the wave number
in vacuum, $c=(\varepsilon_0\mu_0)^{-1/2}$ is the vacuum speed of light
($\varepsilon_0$ and $\mu_0$ are the vacuum permittivity and
permeability, respectively), $\varepsilon_S$ is the dielectric constant of the host surrounding
medium.
The dipole and quadrupole Green's tensors of the medium without particles are
\begin{eqnarray}\label{Green1}
\hat G_{jl}^p\equiv\left\{\left(1+\frac{{
		i}}{k_SR_{jl}}-\frac{1}{k_S^2R_{jl}^2}\right)\right.\hat
U\nonumber\\
+\left.\left(-1-\frac{{
		i}3}{k_SR_{jl}}+\frac{3}{k_S^2R_{jl}^2}\right){\bf
	n}_{lj}{\bf n}_{lj}\right\}\frac{e^{{ i}k_SR_{jl}}}{4\pi R_{jl}}\nonumber\:,\\
\end{eqnarray}
and
\begin{eqnarray}\label{GreenQ}
\hat G^Q_{jl}\equiv\left\{\left(-1-\frac{{
		i}3}{k_SR_{jl}}+\frac{6}{k_S^2R_{jl}^2}+\frac{{
		i}6}{k_S^3R_{jl}^3}\right)\right.\hat
U\nonumber\\
+\left.\left(1+\frac{{ i}6}{k_SR_{jl}}-\frac{15}{k_S^2R_{jl}^2}-\frac{{
		i}15}{k_S^3R_{jl}^3}\right){\bf
	n}_{lj}{\bf n}_{lj}\right\}\frac{{ i} k_S e^{{ i}k_SR_{jl}}}{24\pi R_{jl}}\nonumber\: .\\
\end{eqnarray} 
Here $R_{jl}=|{\bf r}_j-{\bf r}_l|$, ${\bf
	n}_{lj}=({\bf r}_j-{\bf r}_l)/|{\bf r}_j-{\bf r}_l|$ is the unit
vector directed from ${\bf r}_l$ to ${\bf r}_j$, ${\bf n}_{lj}{\bf
	n}_{lj}$ is the dyadic product, $k_S=k_0\sqrt{\varepsilon_S}$,
$\hat U$ is the $3\times3$ unit tensor, $i$ is the imaginary
unit, $\hat Q$ matrix is traceless and symmetric. The vectors
\begin{equation}
{\bf g}_{jl}=\frac{e^{{i}k_S R_{jl}}}{4\pi R_{jl}}\left(\frac{{i}k_S}{R_{jl}}-\frac{1}{R_{jl}^2}\right){\bf R}_{jl}\:
\end{equation}
and 
\begin{equation}
{\bf q}_{jl}=\frac{k_S^2 e^{{i}k_S R_{jl}}}{24\pi R_{jl}^2}\left(1+\frac{3i}{k_S R_{jl}}-\frac{3}{k_S^2R_{jl}^2}\right){\bf R}_{jl}\:.
\end{equation}

\subsection{Coupling in the infinite array}

In the previous works [8,9], it has been shown that, in infinite periodic 2D array with single nanoparticle in the elementary cell, ED does not couple to either EQ [8] or MD [9], and all nanoparticles of the array have the same induced ED, MD, and EQ moments at the normal incidence of external light waves. Here, however, we show that the coupling of EQ and MD is realized in such infinite arrays.

In all derivations below, we consider monochromatic x-polarized incident light with fields $(E_x, H_y, 0)$ and the time dependence $\exp(-i\omega t)$. We also work in the approximations of spherical particles for which ${\bf p}=(p_0\hat x +0\hat y+0\hat z)$, ${\bf m}=(0\hat x + m_0 \hat y + 0\hat z)$, and $\hat Q=Q_0(\hat x\hat z+\hat z\hat x)$, where $\hat x$, $\hat y$ and $\hat z$ are the unit vectors of the Cartesian coordinate system. At these conditions the general system of equations (\ref{E_pmQ}) is written (after straightforward transformations) in the following form
\begin{eqnarray}\label{E_pmQ2}
p_0&=&\frac{\alpha_p}{\varepsilon_0}[\varepsilon_0{ E}_x({\bf r}_0)+ S_1 p_0]\:,\nonumber\\
{m}_0&=&\alpha_m[H_y({\bf
	r}_0)+S_2{ m}_0 +\frac{k_0c}{i}S_3Q_0]\:,\\
{Q}_0&=&\frac{\alpha_q}{2\varepsilon_0}[{\varepsilon_0ik_SE_x({\bf r}_0)}+S_4Q_0+\frac{ik_0}{c}S_5m_0]\:,\nonumber
\end{eqnarray}
where ${\bf r}_0$ is the position of arbitrary particle with number of zero in the array (below we will consider that this particle is located in the origin of Cartesian coordinate system), the polarizabilities $\alpha_p$, $\alpha_m$, and $\alpha_q$ are expressed through the scattering coefficients $a_1$, $b_1$, and $a_2$ of Mie theory [8,9]
$$\alpha_p=i\frac{6\pi\varepsilon_0\varepsilon_S}{k_S^3}a_1,\quad\alpha_m=i\frac{6\pi}{k_S^3}b_1,\quad\alpha_q=i\frac{120\pi\varepsilon_0\varepsilon_S}{k_S^5}a_2,$$
\begin{eqnarray}
S_1&\equiv& k_0^2\sum_{l\neq0}G^p_{xx}(0,{\bf r}_l)=\frac{k_0^2}{4\pi}\sum\limits_{l\ne0}\frac{e^{ik_Sr_l}}{r_l}\left(1+\frac{i}{k_Sr_l}\right.\nonumber\\
&&\left.-\frac{1}{k_S^2r_l^2}-\frac{x_l^2}{r_l^2}-\frac{i3x_l^2}{k_Sr_l^3}+\frac{3x_l^2}{k_S^2r_l^4}\right)\:,
\end{eqnarray}
\begin{eqnarray}
S_2&\equiv& k_S^2\sum_{l\neq0}G^p_{yy}(0,{\bf r}_l)=\frac{k_S^2}{4\pi}\sum\limits_{l\ne0}\frac{e^{ik_Sr_l}}{r_l}\left(1+\frac{i}{k_Sr_l}\right.\nonumber\\
&&\left.-\frac{1}{k_S^2r_l^2}-\frac{y_l^2}{r_l^2}-\frac{i3y_l^2}{k_Sr_l^3}+\frac{3y_l^2}{k_S^2r_l^4}\right)\:,
\end{eqnarray}
\begin{eqnarray}
S_3&\equiv& \frac{1}{Q_0}\sum_{l\neq0}[{\bf q}_{0l}\times(\hat Q^l{\bf
	n}_{l0})]_y\nonumber\\
&=&\sum\limits_{l\ne0}(-x_l^2)\frac{k_S^2e^{ik_Sr_l}}{24\pi r_l^3}\left(1+\frac{3i}{k_Sr_l}-\frac{3}{k_S^2r_l^2}\right)\:,\nonumber\\
\end{eqnarray}
and \cite{Evlyukhin2012}
\begin{eqnarray}
S_4&=&
\frac{ik_0^2k_S}{24\pi}\sum\limits_{l\ne0}\frac{e^{ik_Sr_l}}{r_l^2}\left(-2-i\frac{6+k_S^2x_l^2}{k_Sr_l}\right.\nonumber\\
&&+\frac{12+7k_S^2x_l^2}{k_S^2r_l^2}+i\frac{12+27k_S^2x_l^2}{k_S^3r_l^3}\nonumber\\
&&\left.-\frac{60x_l^2}{k_S^2r_l^4}-\frac{i60x_l^2}{k_S^3r_l^5}\right)\:,
\end{eqnarray}
and
\begin{eqnarray}
S_5&=&6S_3=
\sum\limits_{l\ne0}\frac{x_l^2k_S^2e^{ik_Sr_l}}{4\pi r_l^3}\left(-1-\frac{3i}{k_Sr_l}+\frac{3}{k_S^2r_l^2}\right)\:.\nonumber\\
\end{eqnarray}
Most importantly, the coefficients $S_3$ and $S_5$ in (\ref{E_pmQ2}) are not equal to zero for infinite arrays providing coupling between MD ($m_0$) and EQ ($Q_0$) moments. 

Solving of the system (\ref{E_pmQ2}) one can find effective polarizabilities of the particles in the array. These polarizabilities are determined by the expressions: $\alpha^{\text{eff}}_p=p_0/E_x$; $\alpha^{\text{eff/coup}}_m=m_0/H_y$; $\alpha^{\text{eff/coup}}_q=2Q_0/(ik_SE_x)$.
Thus we obtain 
\begin{equation}\label{Am}
\frac{1}{\alpha_m^{\text{eff/coup}}}=\frac{1-S_5\alpha_m^{\text{eff}}S_3\alpha_q^{\text{eff}}k_0^2/(2\varepsilon_0)}{\alpha_m^{\text{eff}}[1+S_3\alpha_q^{\text{eff}}k_0^2/(2\varepsilon_0)]}\:,
\end{equation}
\begin{equation}\label{Aq}
\frac{1}{\alpha_q^{\text{eff/coup}}}=\frac{1-S_5\alpha_m^{\text{eff}}S_3\alpha_q^{\text{eff}}k_0^2/(2\varepsilon_0)}{\alpha_q^{\text{eff}}[1+S_5\alpha_m^{\text{eff}}]}\:,
\end{equation}
\begin{equation}\label{Ap}
\frac{1}{\alpha_p^{\text{eff}}}=\frac{1}{\alpha_p}-\frac{S_1}{\varepsilon_0}\:,
\end{equation}
where
$$
\frac{1}{\alpha_m^{\text{eff}}}=\frac{1}{\alpha_m}-S_2\:,\quad \frac{1}{\alpha_q^{\text{eff}}}=\frac{1}{\alpha_q}-\frac{S_4}{2\varepsilon_0}\:.
$$

\subsection{Reflection and transmission}

Estimation of the reflection and transmission coefficients can be
obtained if we consider total electric field in the
far-field region for $z<0$ and $z>0$. 
\begin{equation}
{\bf E}={\bf E}_{0}+{\bf E}_p+{\bf E}_m+{\bf E}_q
\end{equation}
The electric fields generated by ED, MD, and EQ of nanoparticles are 
\begin{equation}
{\bf E}_p=\frac{k_0^2}{\varepsilon_0}
E_x\alpha_p^{\text{eff}}(G_{xx}^r,0,0)
\end{equation}
\begin{equation}
{\bf E}_m=-\frac{ik_0}{c\varepsilon_0}
H_y\alpha_m^{\text{eff/coup}}(g^r_z,0,0),
\end{equation}
\begin{equation}
{\bf
	E}_q=\frac{k_0^2}{\varepsilon_0}\frac{ik_SE_x}{2}\alpha_q^{\text{eff/coup}}(G_x^{Q,r},0,0),
\end{equation}
respectively.
For the case of wavelength larger than the array lattice periods, non-zero total electric field component $E_x^f$ in the far field approximation is 
 \begin{eqnarray}
E_x^f&=&E_x\left[e^{ik_Sz}+\frac{k_0^2}{\varepsilon_0}\:
\alpha_p^{\text{eff}}G_{xx}^r-ik_S\alpha_m^{\text{eff/coup}}g_{z}^r\right.\nonumber\\
&& +\left.\frac{k_0^2}{\varepsilon_0}\:
\alpha_q^{\text{eff/coup}}\:\frac{ik_S}{2}\:G_{x}^{Q,r}\right]
\end{eqnarray}
where [9]
$$
G_{xx}^r=\frac{i}{2S_Lk_S}e^{\mp ik_Sz}\:,\quad g_z^r=\pm\frac{1}{2S_L}e^{\mp ik_Sz}
$$
and (the calculation method as in [9])
$$
G_x^{Q,r}=\mp\frac{1}{12S_L}e^{\mp ik_Sz}\:,
$$
$S_L$ is the area of the lattice unit cell and the upper sing
corresponds to $z<0$ and the lower sing for the case when $z>0$. In
this approach the reflection (for electric field) and transmission
coefficients are (compare with [9]) 
\begin{equation}\label{rrr}
r=\frac{ik_S}{2S_L}\left[\frac{1}{\varepsilon_0\varepsilon_S}\alpha_p^{\text{eff}}-\alpha_m^{\text{eff/coup}}-\frac{k_0^2}{12\varepsilon_0}\:\alpha_q^{\text{eff/coup}}\right]\:,
\end{equation}
\begin{equation}\label{ttt}
t=1+\frac{ik_S}{2S_L}\left[\frac{1}{\varepsilon_0\varepsilon_S}\alpha_p^{\text{eff}}+\alpha_m^{\text{eff/coup}}+\frac{k_0^2}{12\varepsilon_0}\:\alpha_q^{\text{eff/coup}}\right]\:.
\end{equation}
The intensity reflection $R_0$ and transmission $T_0$ coefficients are 
$$
R_0=|r|^2\:,\quad T_0=|t|^2\:.
$$ 

Calculations of the contributions $\alpha_m^{\text{eff/coup}}$ and ${k_0^2}\:\alpha_q^{\text{eff/coup}}/({12\varepsilon_0}) $ show that their magnitudes at resonance are comparable to $\alpha_p^{\text{eff}}/({\varepsilon_0\varepsilon_S})$ (Fig. 5a,b). One can also note that without coupling, $\alpha^{\text{eff}}_m$ does not possess any resonant features. In the case of coupling, both EQ and MD lattice resonances are excited spectrally close to Rayleigh anomaly, while without coupling (or in the case of zero MD response) EQ lattice resonance is shifted further away from RA similar to well-known ED lattice resonance shifts.
In Fig. 5d,e, we demonstrate an enlarged view of spectra for spheres with $R = 100$ nm, and we show that reflectance minimum and transmittance maximum almost coincide with the maximum of absorbance, where EQ-MD lattice resonance is excited. Comparison of analytical calculations with Eqs. (\ref{rrr}), (\ref{ttt}) with numerical modeling performed with CST Microwave Studio demonstrates a very good quantitative agreement (Fig. 5d,e). It also shows that taking into account EQ-MD coupling is critical for obtaining the correct profile and position of Kerker effect (Fig. 5d). The phase change of transmittance coefficient experiences a rise in the proximity of lattice resonance (Fig. 5c). Thus, we observe resonant Kerker effect enabled by destructive interference between the ED resonance of a single particle in the array and the EQ-MD lattice resonance of the particle with non-resonant contribution of EQ and MD moments.

\begin{figure*}
	\begin{center}
		\includegraphics[scale=.100,height=10pc,width=13pc,keepaspectratio]{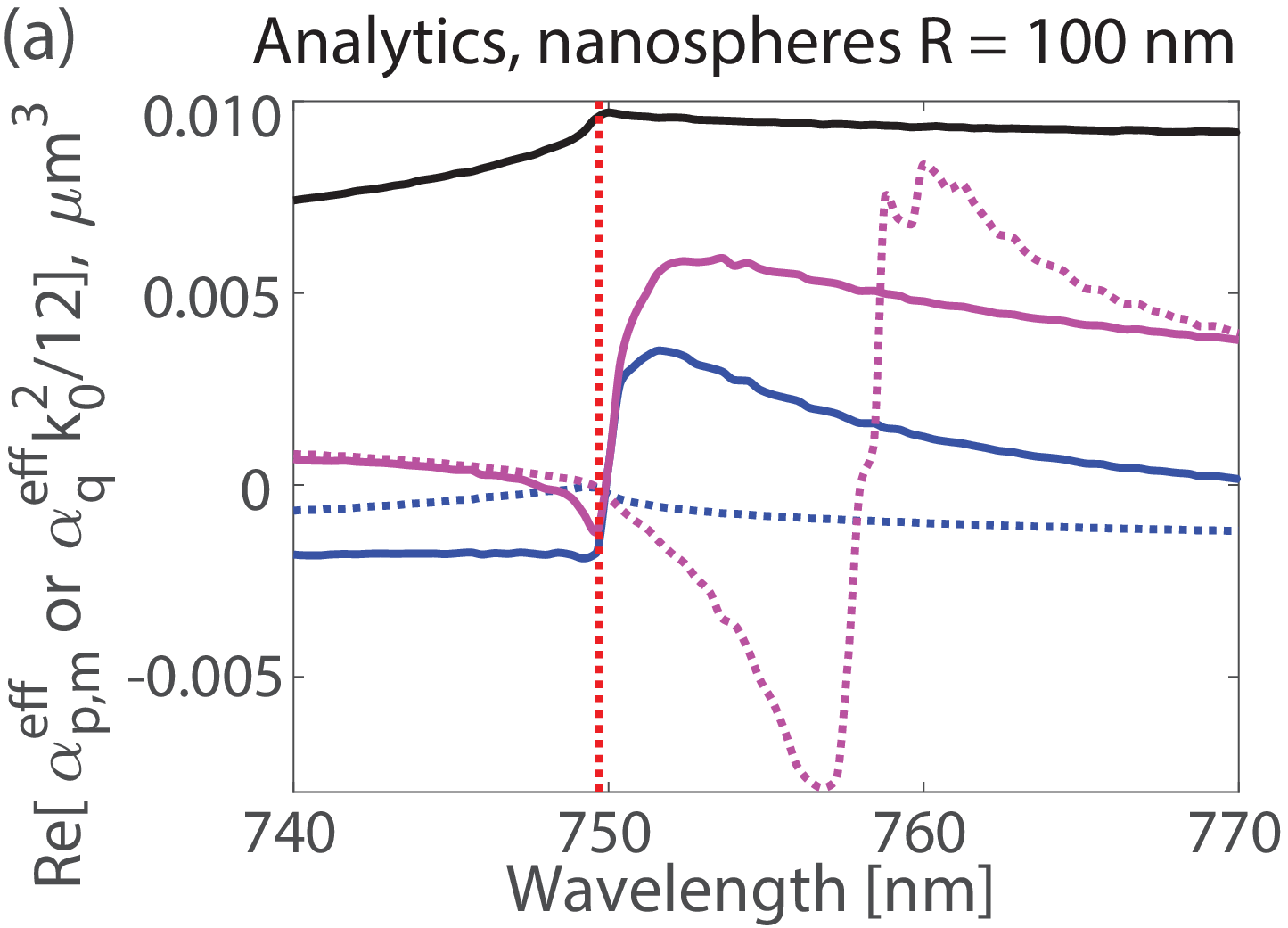}
		\includegraphics[scale=.100,height=10pc,width=13pc,keepaspectratio]{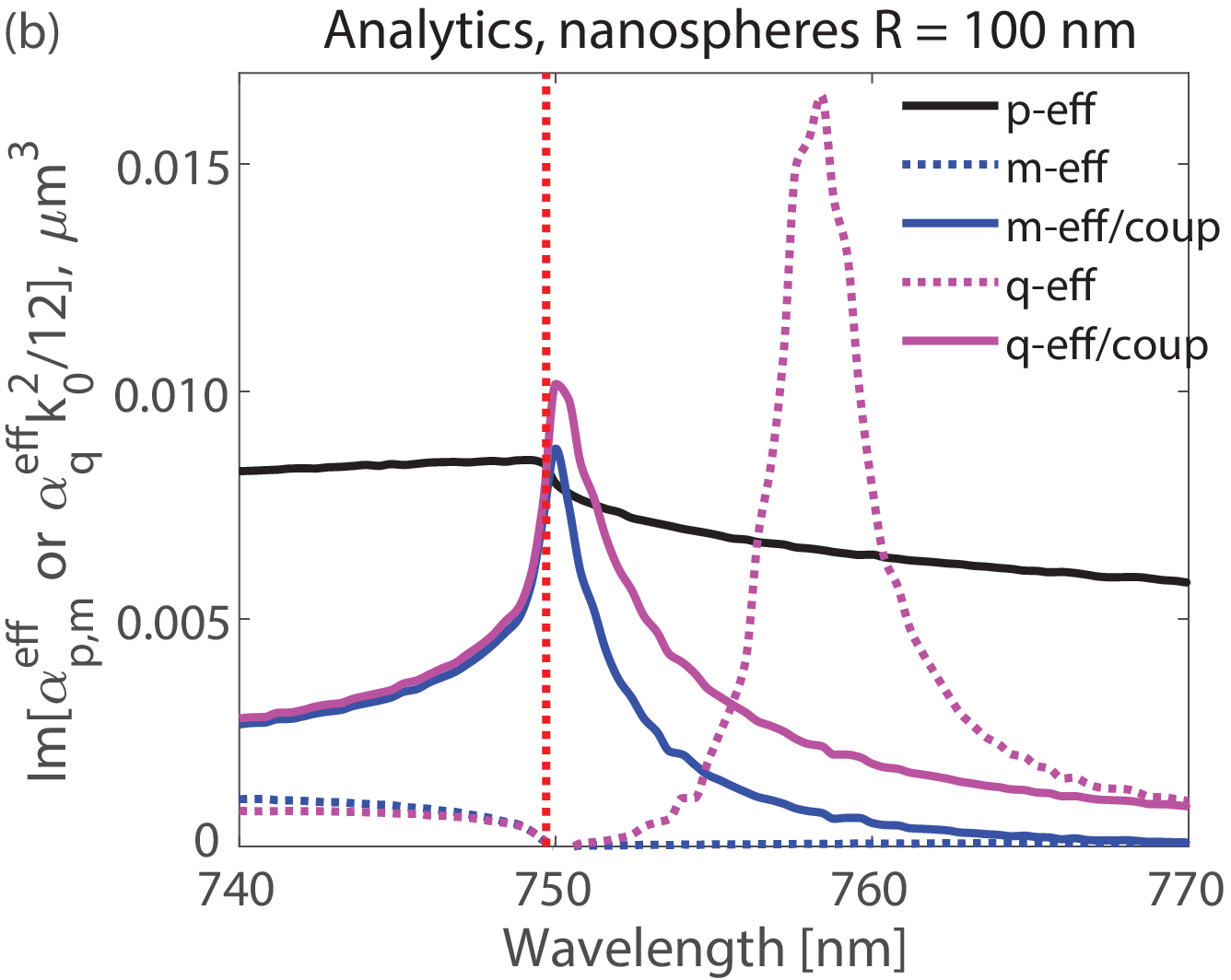}
		\includegraphics[scale=.100,height=10pc,width=13pc,keepaspectratio]{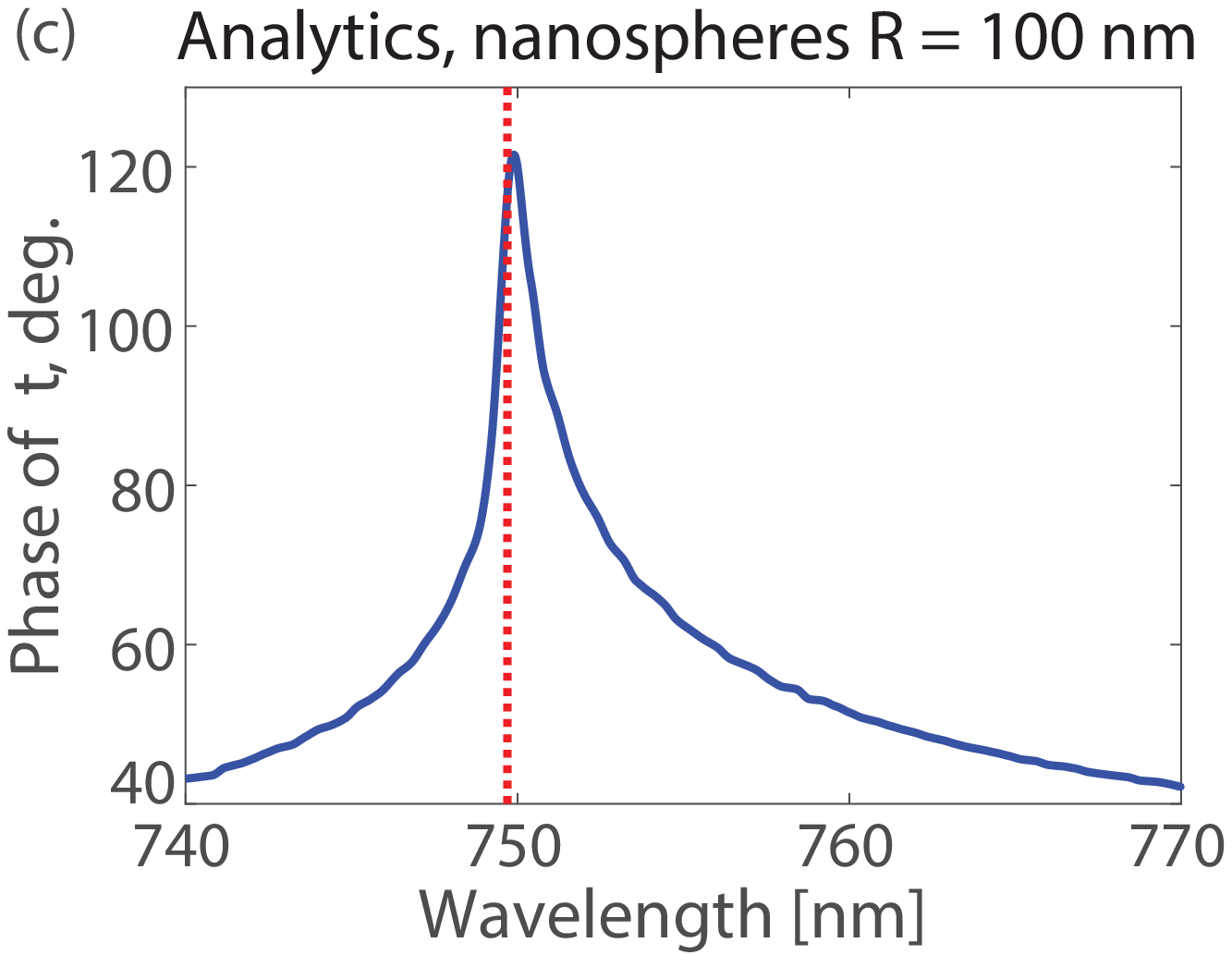}
		
		\includegraphics[scale=.100,width=16pc,height=12pc,keepaspectratio]{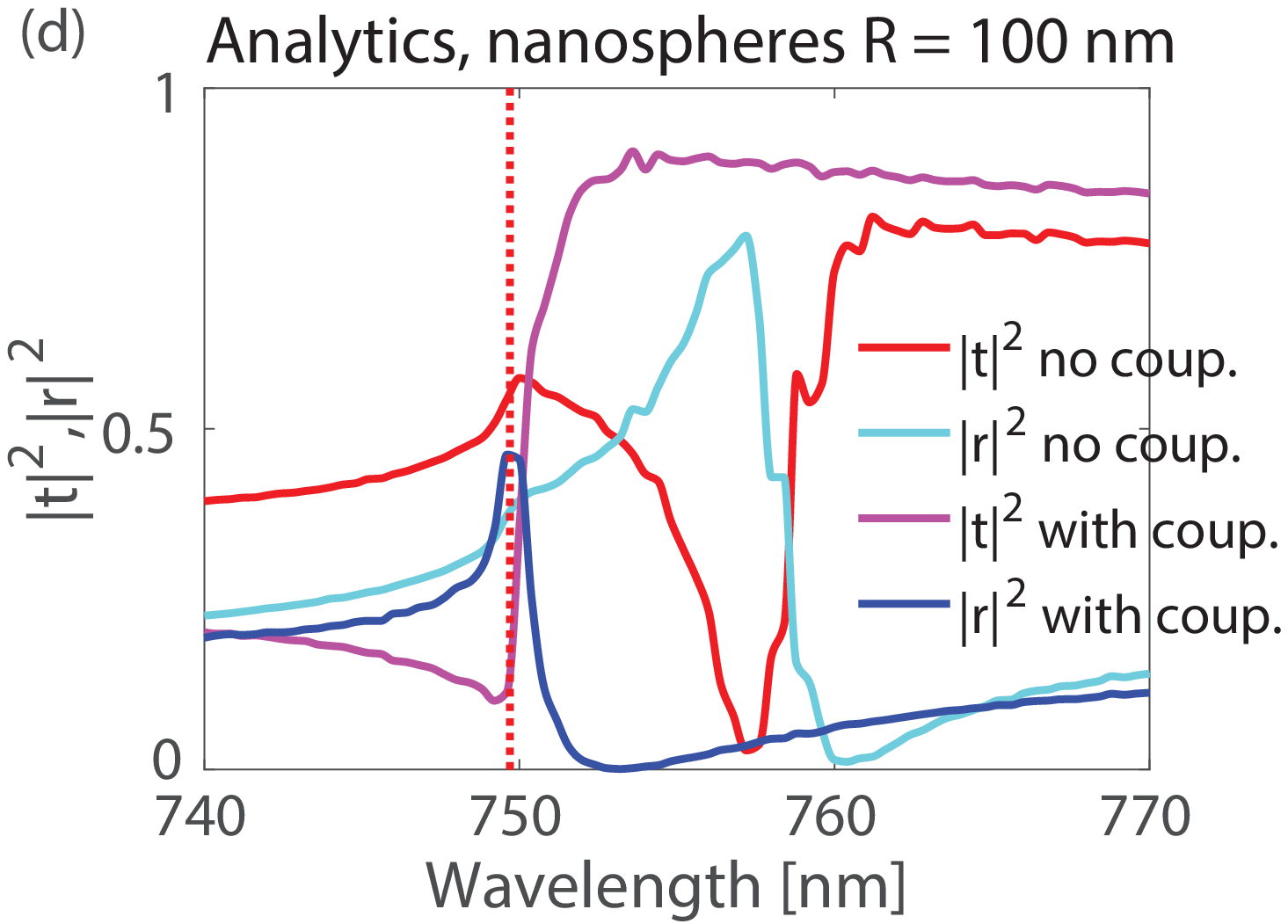}
		\includegraphics[scale=.100,width=14pc,height=11pc,keepaspectratio]{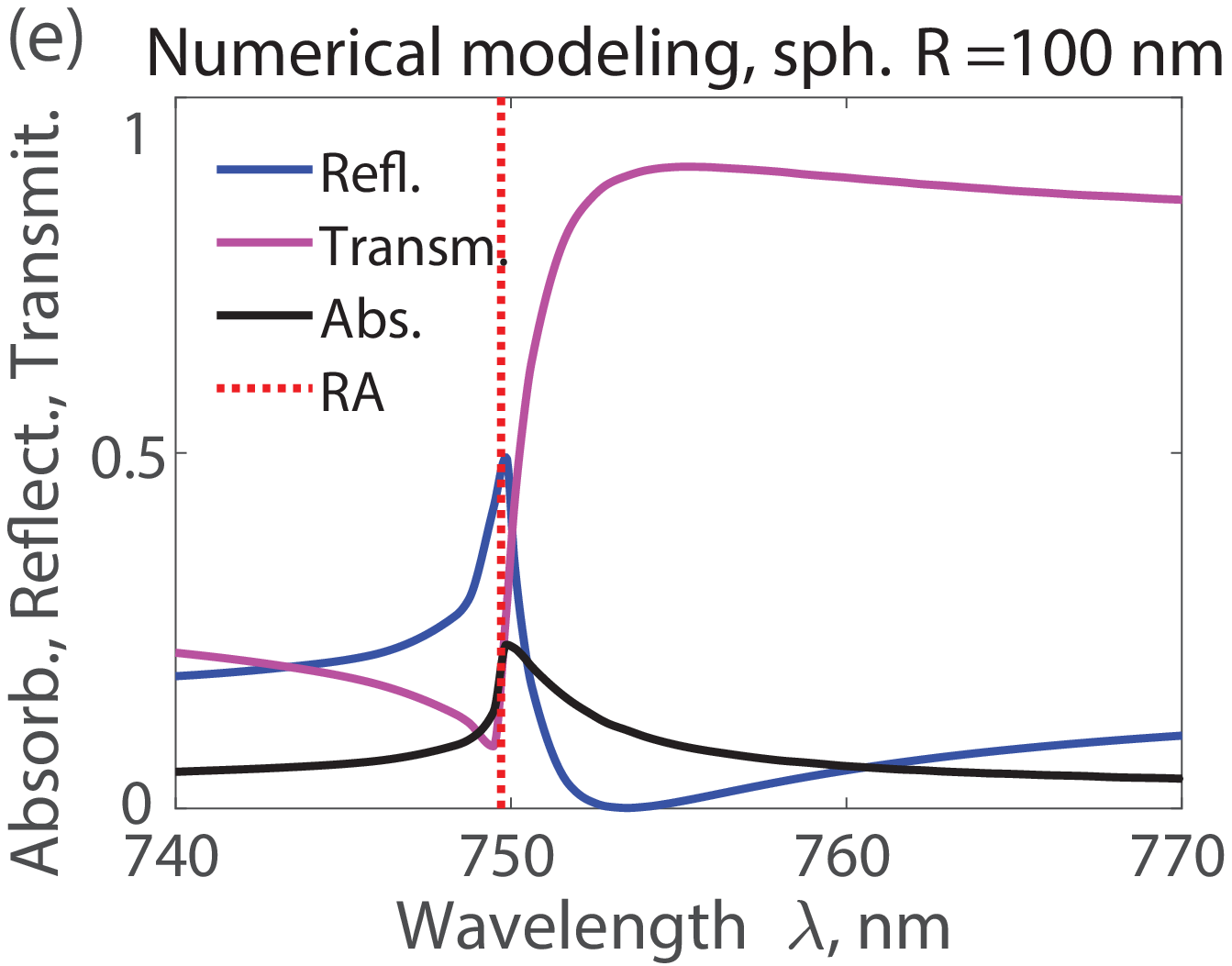}
	\end{center}
	\caption{\label{F5} Nanosphere array. (a) Real and (b) imaginary parts of $\alpha_m^{\text{eff/coup}}$ and
	 ${k_0^2}\:\alpha_q^{\text{eff/coup}}/({12\varepsilon_0})$ with and without coupling of EQ and MD (see Eqs. (\ref{Am}) and (\ref{Aq})). Legend is the same for both (a) and (b) panels. Polarizability of the ED is not affected by other multipoles (see Eq. (\ref{Ap})). RA line denotes the wavelength of Rayleigh anomaly. (c) Change of phase of transmittance coefficient $t$ calculated by Eq. (\ref{ttt}). (d) Analytical calculations of reflectance and transmittance for the cases without EQ and MD coupling 'no coup.' and with EQ and MD coupling 'with coup.' (e) Reflectance, transmittance, and absorbance calculated in numerical modeling with CST Microwave Studio. One can see that the reflectance minimum and transmittance maximum almost coincide with the maximum of absorbance, where EQ lattice resonance is excited. The nanoparticle array with spheres of $R = 100$ nm is in the homogeneous environment with $n = n_s = 1.47$. Periods are $p_x = 510$ nm and $p_y = 250$ nm. }
\end{figure*}

\subsection{Nanodisk array}

In contrast to spheres, disks' polarizability is anisotropic, and excitation of different modes depends on light illumination conditions. We calculate a multipole decomposition [50] of scattering cross-sections for single disks under normal light incidence (Fig. 6). The results of the calculations for the thick disk with $H = 100$ nm show that total scattering cross-sections have a broad ED-resonant peak at wavelengths 800-900 nm and a narrow EQ-resonant peak at 630 nm wavelength as well as the minor MD peak at 650 nm. For the disk with the smallest height $H = 50$ nm, only ED resonance is present, and EQ and MD contributions are very weak (Fig. S5a in Supporting Information). Nevertheless, for the disks with $H = 50$ nm, transmittance dip is at wavelength 750 nm (Fig. 6d), and the absorbance profile has both broad and narrow peaks (Fig. 6e). For the disks with $H = 100$ nm, there is also a broad dip in transmittance. In contrast, in absorbance, a narrow feature is the strongest in comparison to the spheres of all sizes and disks with $H = 50$ nm. Thus, even though the magnitudes of EQ and MD moments are not pronounced in comparison to ED resonance, their excitation is responsible for the lattice resonance. However, its spectral width is very narrow, which corresponds to extremely long propagation length of the excited lattice mode [6]. Similar to nanospheres array, lattices of nanodisks demonstrate a suppressed reflection at the EQ-MD lattice resonance (Figs. 6c,7), which corresponds to destructive interference of EQ, MD, and ED scattering in a backward direction mimicking the resonant Kerker effect. The reflectance and transmittance profiles are similar to those of nanospheres (compare Figs. 7 and 5e).

\begin{figure*}
	\begin{center}
		\includegraphics[scale=.100,width=13pc,height=10pc,keepaspectratio]{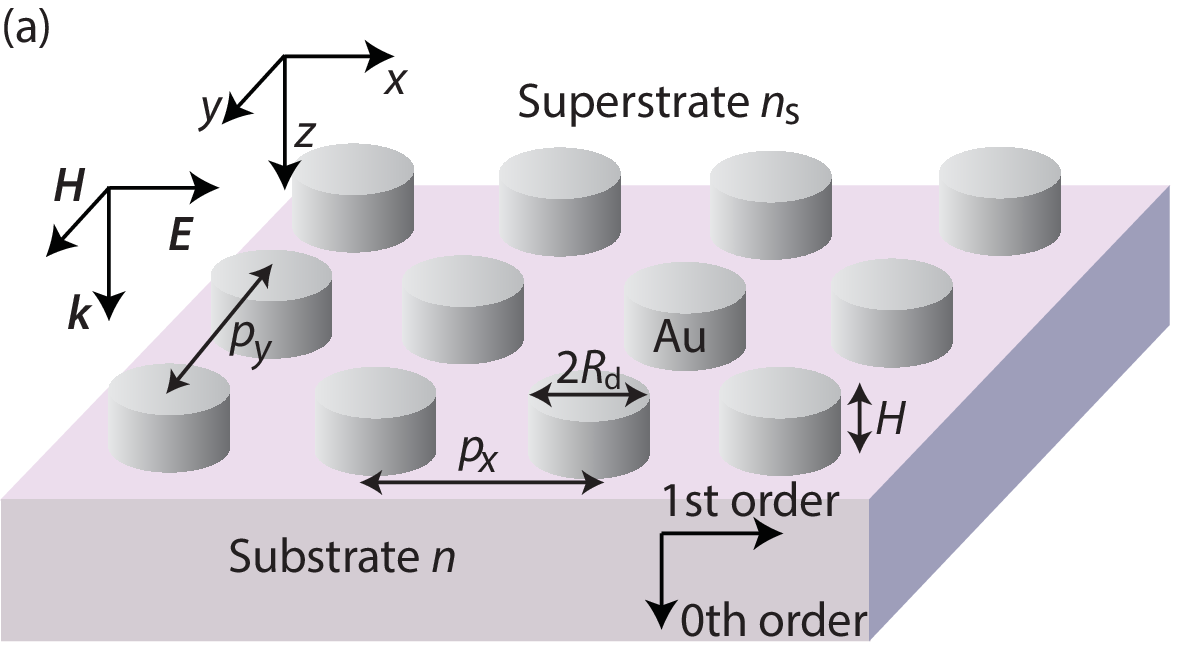}
		\includegraphics[scale=.100,width=13pc,height=10pc,keepaspectratio]{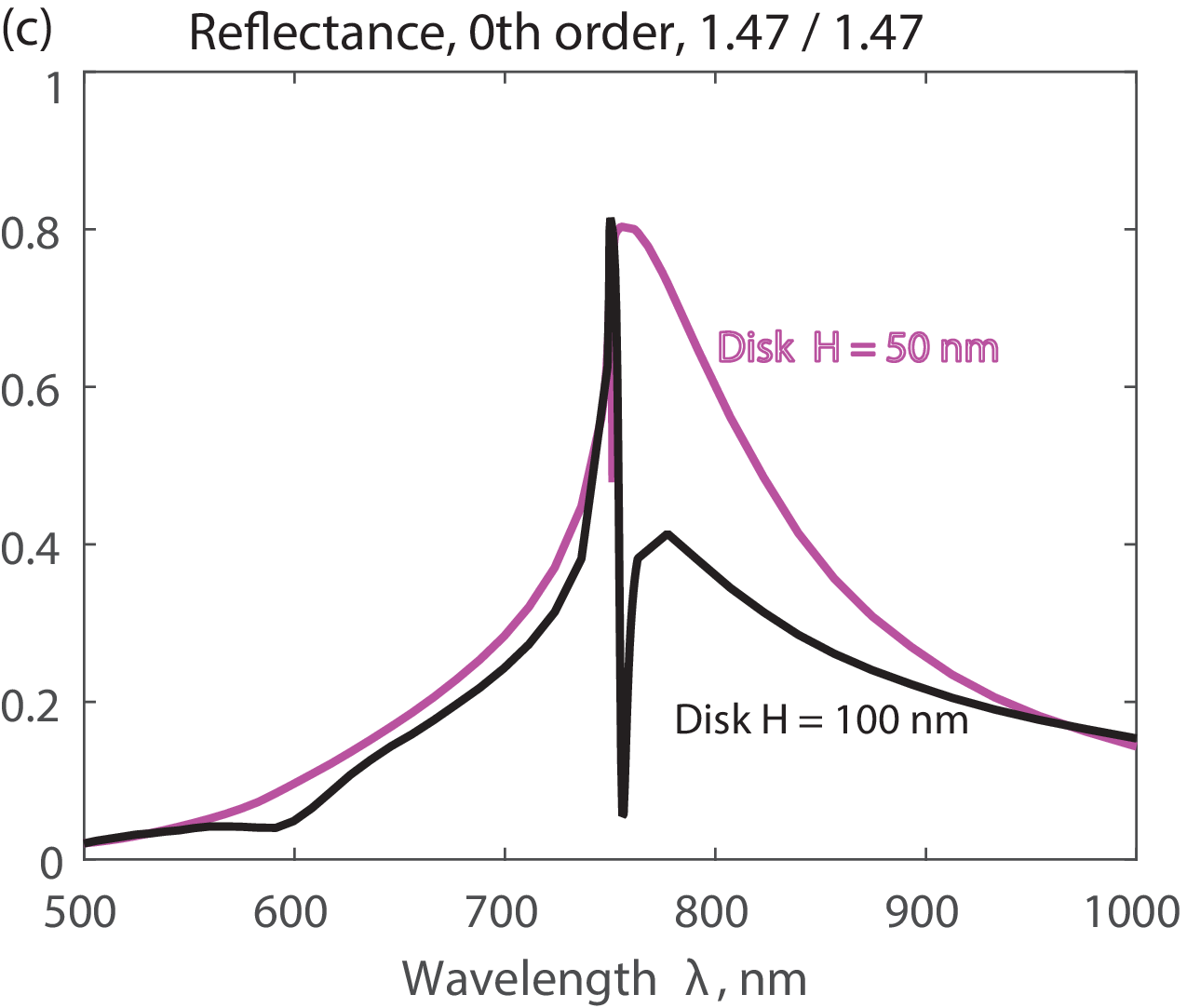}
		\includegraphics[scale=.100,width=13pc,height=10pc,keepaspectratio]{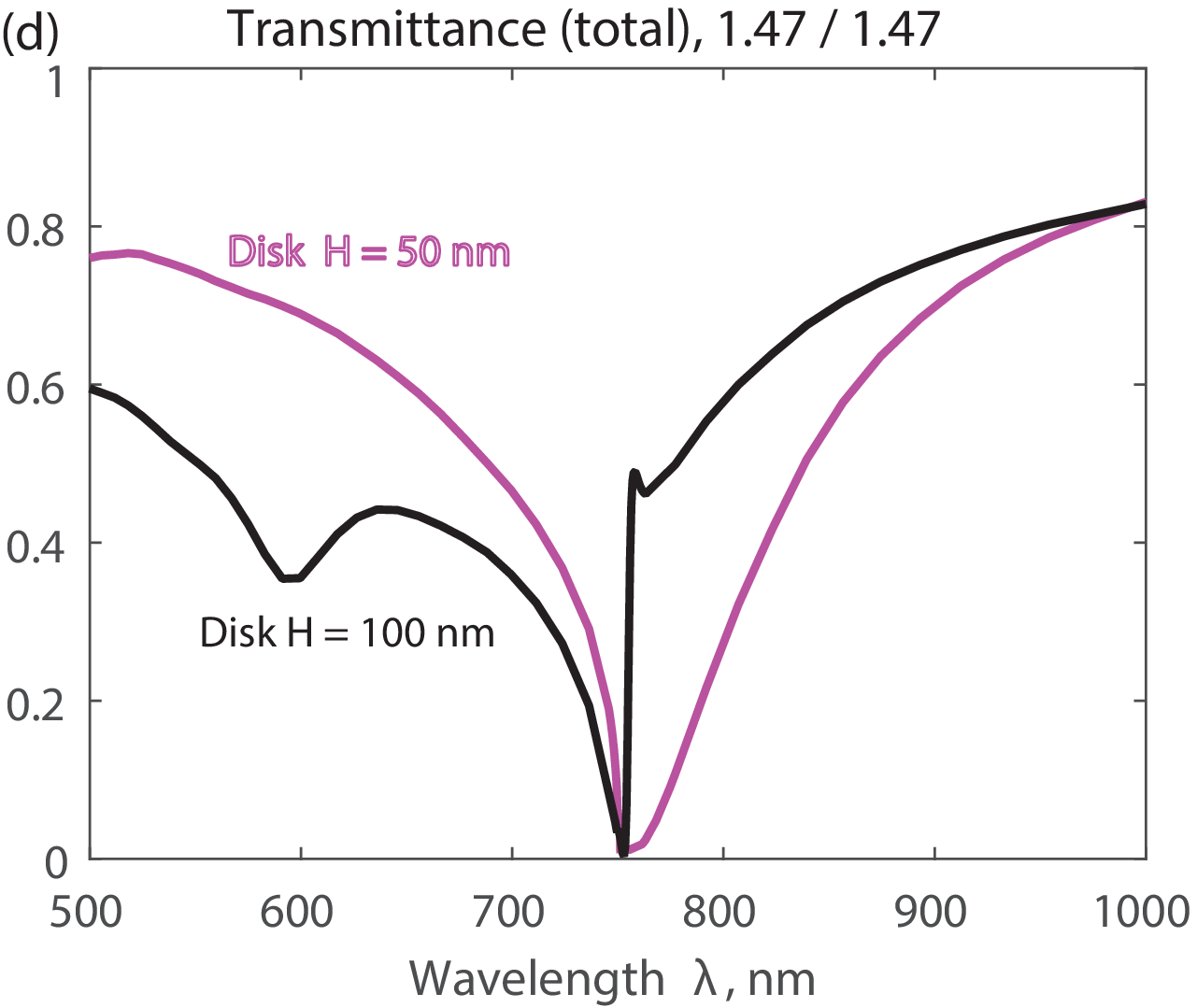}	
		
		\includegraphics[scale=.100,width=14pc,height=11pc,keepaspectratio]{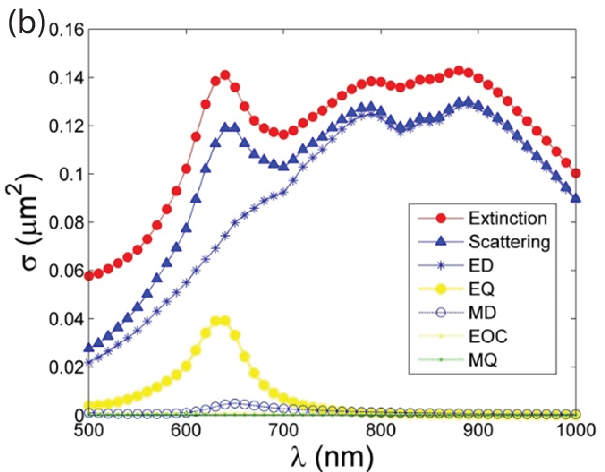}	
		\includegraphics[scale=.100,width=28pc,height=11pc,keepaspectratio]{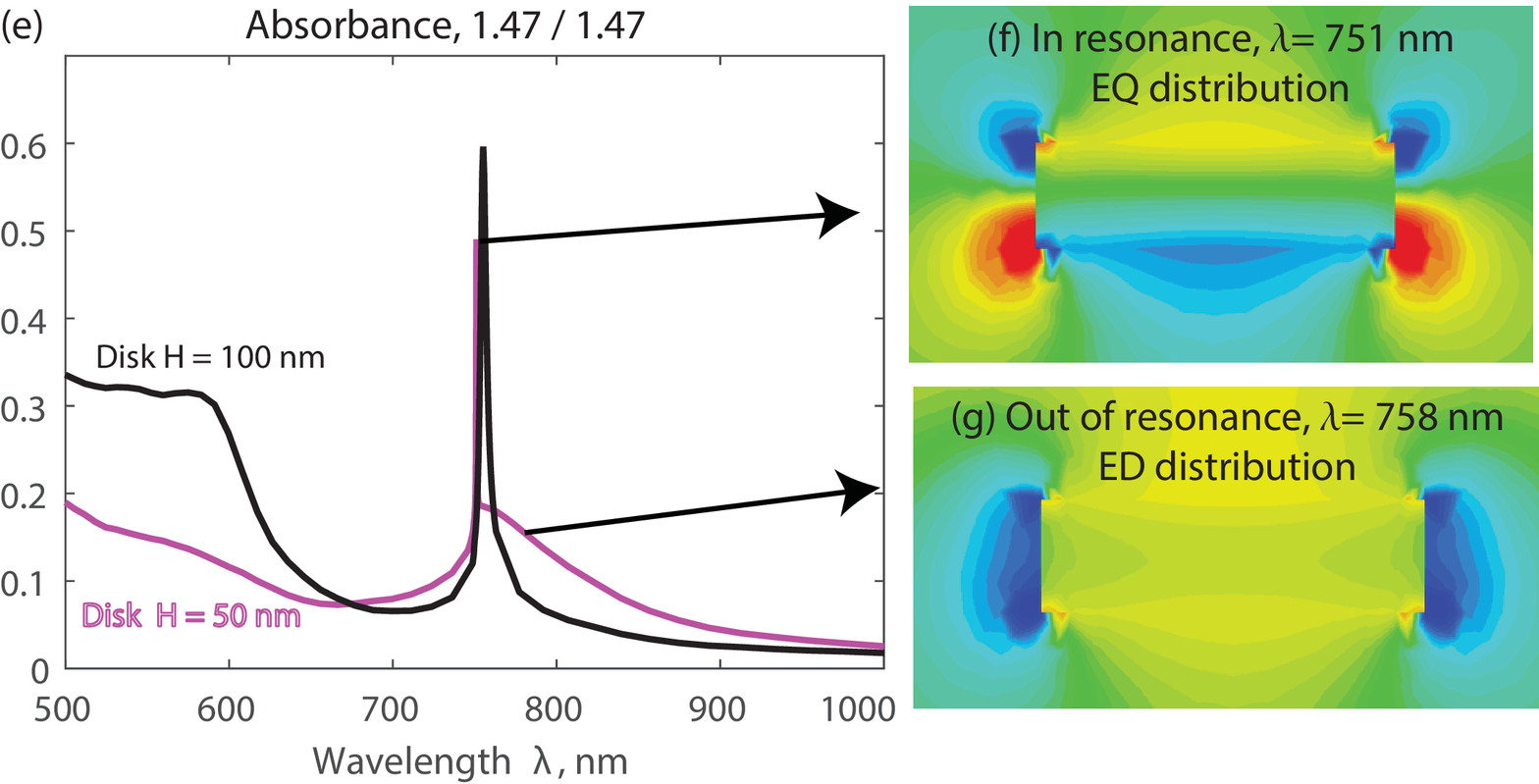}
	\end{center}
	\caption{\label{F6} (a) Periodic array of the nanodisks with radius $R_d$ and height $H$. (b) Total extinction/scattering cross-sections and multipole decomposition of the nanodisks with $R_d = 85$ nm and $H = 100$ nm in a homogeneous environment with $n = n_s = 1.47$ calculated using a discrete dipole-quadrupole approximation. EOC is electric octupole, and MQ is magnetic quadrupole. (c) Zero order reflection, (d) total transmission, and (e) absorbance of the nanodisk array with $R_d = 85$ nm in the homogeneous environment with $n = n_s = 1.47$. Periods are $p_x = 510$ nm and $p_y = 250$ nm. (f) In-resonance ($\lambda = 751$ nm) and (g) out-of-resonance ($\lambda = 758$ nm) field distributions for the array of nanodisks with $H = 50$ nm (one unit cell, side view). EQ field distribution is well pronounced for the resonant wavelength, and the field distribution drastically changes to ED with the small change of wavelength. See more simulation results in Supporting Information including the case of the non-homogeneous environment, when the refractive indices of the superstrate and substrate are $n_s = 1.47$ and $n = 1.5$, respectively (Figs. S3 and S4). }
\end{figure*}

\begin{figure}
	\begin{center}
		\includegraphics[scale=.100,width=16pc,height=12pc,keepaspectratio]{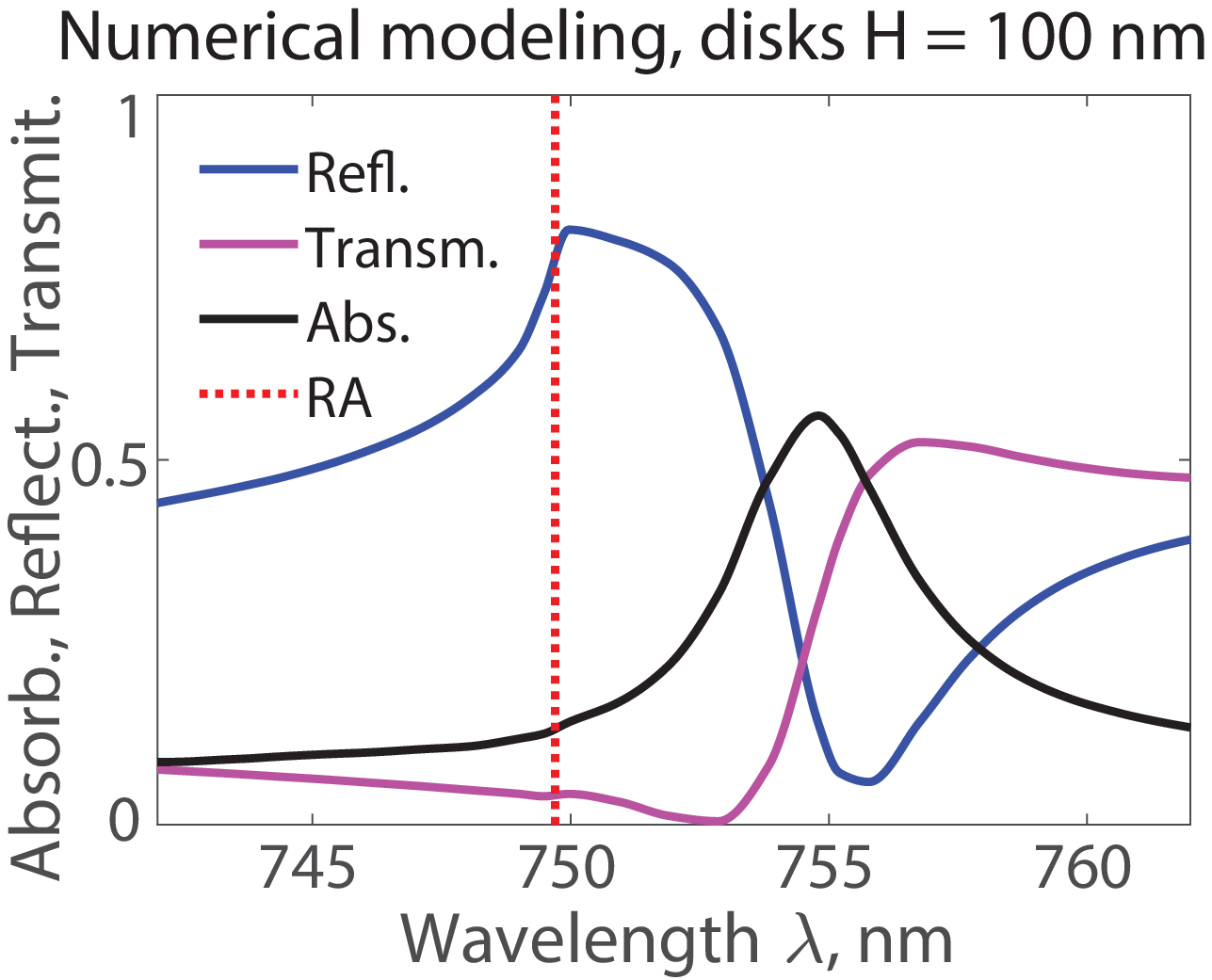}%{S_p_small}%{ReSumR120n1_45Rs660}% 
	\end{center}
	\caption{\label{F7}Reflectance, transmittance, and absorbance of the nanodisk array, $H = 100$ nm, in the homogeneous environment with $n = n_s = 1.47$. Periods are $p_x = 510$ nm and $p_y = 250$ nm. RA line denotes the wavelength of Rayleigh anomaly. One can see that the reflectance and transmittance profiles are similar to those of nanospheres (Fig. 5(e)).}
\end{figure}

Further, we study nanoparticle array in a non-homogeneous environment that is a structure where particles (disks with $R_d = 85$ nm and $H = 100$ nm) are placed on the substrate with varied $n$, while for the superstrate, $n_s = 1.47$ is fixed (Fig. 8). In this case, we also observe the collective resonances, and due to the differences between the diffraction conditions in the substrate and superstrate, the resonant signature in the transmission and absorption spectra is split into two features. One can see that the transmittance spectra include narrow transmission bands between two Rayleigh anomalies that correspond to the substrate and superstrate. Strong absorption in the structure exists in homogeneous surrounding and disappears with increasing of difference between $n$ and $n_s$ (Fig. 8a).

\begin{figure*}
	\begin{center}
		\includegraphics[scale=.100,width=42pc,height=13pc,keepaspectratio]{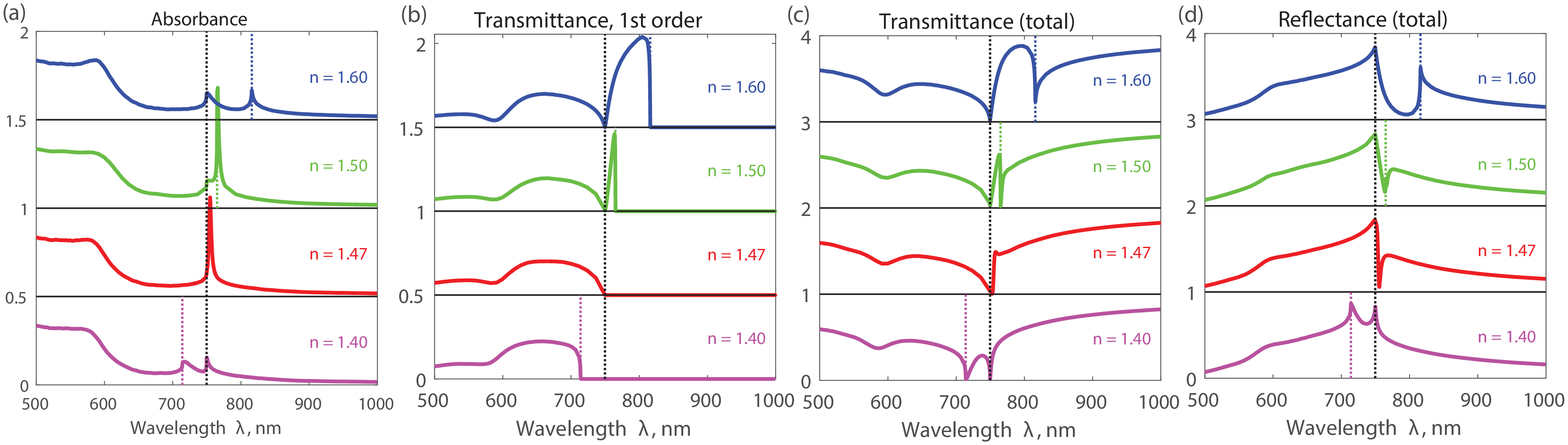}
		\includegraphics[scale=.100,width=39pc,height=9pc,keepaspectratio]{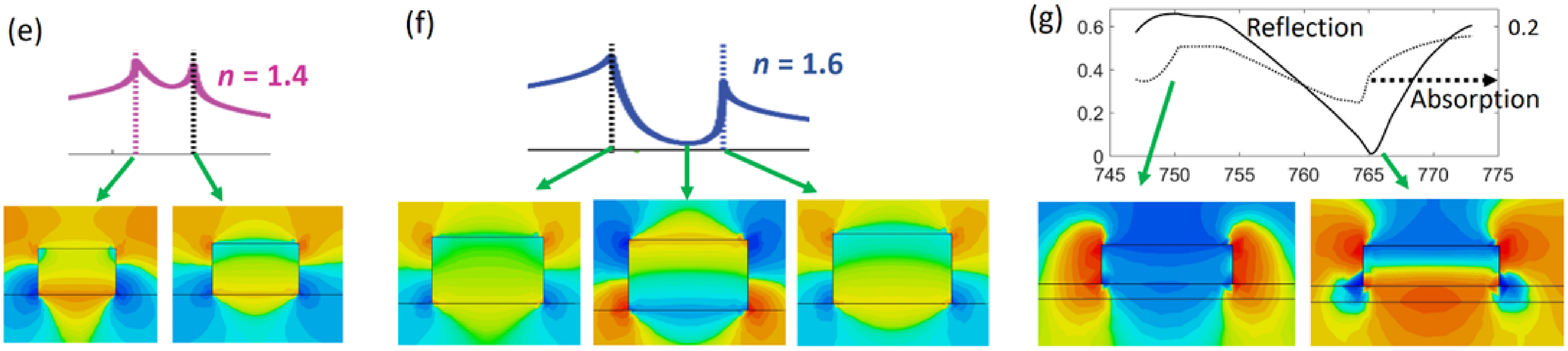}
	\end{center}
	\caption{\label{F8} Lattice resonance in an array of nanodisks when substrate and superstrate have different refractive indices. (a) The absorbance, (b) first order transmission, (c) total transmission, and (d) total reflection of the array of nanodisks ($R_d = 85$ nm and $H = 100$ nm) for different $n$ denoted in the plots. Periods are $p_x = 510$ nm and $p_y = 250$ nm, $n_s = 1.47$. The dotted color lines mark a wavelength where the resonances in the substrate are expected, i.e. $\lambda_{\text{RA}} = np_x$, and the dotted black line corresponds to the resonance in superstrate, i.e. $\lambda_{\text{RA}s} = n_sp_x$. Each plot is shifted by either 0.5 or 1 with respect to the previous one. See more simulation results for this array in Supporting Information (Figs. S8). Similar behavior has been obtained for disks with $H = 50$ nm (Fig. S9) and other periods of the array (not shown here). Analysis of absorption spectra for the other set of parameters ($p_x = 420$ nm and $1.4 < n < 2.15$) is presented in the Supporting Information (Fig. S10). (e)-(g) Field distributions at different parts of the reflectance profiles (top parts): (e) for $n = 1.4$; (f) for $n = 1.6$; and (g) corresponds to the experimental structure in Ref. [30]. Quadrupole field distribution is the most pronounced for near-zero reflectance for $n = 1.6$ and experimental structure in Ref. [30]. }
\end{figure*}

To clarify the reason for the narrow transmittance band in the case of non-homogeneous systems, we analyze the role of EQ resonance excitation at each Rayleigh anomaly in the spectral profiles (Fig. 8e,f). Weak EQ resonances are excited in the cases of $n = 1.4$ and $n = 1.6$ at the exact position of Rayleigh anomalies, but well pronounced EQ is observed for the wavelength where the reflectance is almost zero (Fig. 8f). There, reflections from two EQ and two MD resonances compensate dipole reflection, and the Kerker effect is observed in the region between resonances.

We have performed numerical simulations of the experimental structure in Ref. [30]: disks with $R_d = 85$ nm, $H = 50$ nm, periods are $p_x = 510$ nm and $p_y = 250$ nm, indices $n_s = 1.47$, $n = 1.5$, and 20-nm ITO layer with $n_{\rm ITO} = 1.9$ appears below the disks. At $\lambda\approx n_sp_x\approx750$ nm, both reflection and absorption have local maximum, and field distribution corresponds to the ED resonance (Fig. 8g). At the same time, at $\lambda\approx n p_x\approx765$ nm, both reflection and absorption have minima, reflection is nearly zero, and field distribution corresponds to EQ resonance. This confirms that directional scattering and Kerker effect appear at the wavelength of strong EQ and MD excitations.

In Supporting Information, Fig. S6 schematically shows transmission and reflection in the zeroth and first orders of diffraction and their changes that depend on the incident wavelength with respect to Rayleigh anomaly in the substrate and the superstrate. We calculated the cases with a small difference between the refractive indices of the substrate and the superstrate, in particular, $n_s = 1.47$ and $n = 1.48$. There, the substrate-induced transparency is narrower and stronger in comparison to the case $n = 1.5$ (see Supporting Information, Fig. S7). The smaller the difference between indices the stronger and the narrower the transmittance peak, and it appears between the two Rayleigh anomalies associated with the substrate and superstrate. For equal indices, the transmittance peak vanishes. 
The substrate-induced transparency effect is essentially different from the effects of the substrate observed before in [4,51-54]. In contrast to the previous works, where the large difference between medium indices causes smearing out of the lattice resonances, in our work, the difference is small, resonance conditions in each medium are similar, and because of the effective "broadening", the lattice resonances are more pronounced. In the experimental work [30], substrate and index-matching superstrate refractive indices are the fabrication process, which is associated with fabrication process, and particularly this difference has given a rise to well-pronounce resonances features.

\section{Conclusions }

It is well known that dipole coupling in one- and two-dimensional plasmonic nanoparticle arrays can produce narrow collective plasmon resonances in light transmission spectra, and the wavelengths are determined by the array periods. In the electric dipole approximation, the collective lattice resonances involve only dipole moments of the nanoparticles oriented perpendicular to the lattice wave propagation. We have studied transmittance, reflectance, and absorbance of periodic nanoparticle arrays in parallel polarization, where only EQ and MD moments are involved in lattice resonances, but the ED moment is broad and out of its lattice resonance. It has been shown that these arrays can support excitation of lattice mode resonances in both homogeneous and non-homogeneous environments. The effect cannot be described using only electric dipole approximations, and higher multipoles coupling between nanoparticles are necessary to explain pronounced resonant features. Even for very small, but non-zero, EQ and MD moments, their lattice resonances are strong enough to provide destructive interference with the electric dipole. This results in zero reflectance associated with a resonant Kerker effect where the generalized condition of directional scattering is satisfied. The most importantly, we have found that in the infinite periodic nanoparticle lattice even in the homogeneous environment, EQ and MD moments of nanoparticles are coupled and affect each other resonant contributions.

In the experimental realization with finite size particles arrays, the excitation of the lattice mode is difficult. Nevertheless, the resonance broadening is possible in the case of the non-homogeneous environment. Owing to an excitation of collective multipole resonances in the nanoparticle arrays in the non-homogeneous environment, e.g. with different substrate and superstrate indices, a narrow transparency band can appear between two Rayleigh anomalies that correspond to disappearing of the reflection and transmission first diffraction orders in the substrate and superstrate, respectively. For the small difference between indices, strong substrate-induced transparency can be achieved along with a narrow peak in absorbance profile. 

We have concluded that experimentally observed resonances in parallel polarization in Ref. [30] are the result of the presence of substrate and superstrate of different refractive indices and excitation of resonant lattice modes supported by the multipole coupling. In terms of practical applications, because of the high spectral sensitivity of collective resonances to the environment, these lattice resonances can be used for sensing.

 $\:$
{\bf Acknowledgments}
\noindent 

%\begin{acknowledgement}

The authors acknowledge financial support from the Deutsche
Forschungsgemeinschaft (Germany), the project EV 220/2-1. The numerical study has been supported by the Russian Science Foundation (Russian Federation), the project 16-12-10287.

%\end{acknowledgement}

%\begin{suppinfo}

%Additional information is provided on transmittance, reflectance, and absorbance profiles for different parameters of the %nanoparticle array. This material is available free of charge via the Internet at http://pubs.acs.org.

%\end{suppinfo}

\end{document}